\documentclass[aps,prapplied,twocolumn,superscriptaddress,floatfix, amsmath,amssymb,showpacs,longbibliography]{revtex4-1}
\usepackage{graphicx}
\usepackage[colorlinks,allcolors=blue]{hyperref}
\usepackage{color}
\usepackage{soul}
\usepackage{ulem}

\begin{document}

\title{Floquet-engineered enhancement of coherence times in a driven fluxonium qubit}
\author{Pranav S. Mundada}\thanks{These authors contributed equally to this work.}
\affiliation{Department of Electrical Engineering, Princeton University, Princeton, New Jersey 08540, USA}
\author{Andr\'as Gyenis}\thanks{These authors contributed equally to this work.}
\affiliation{Department of Electrical Engineering, Princeton University, Princeton, New Jersey 08540, USA}
\author{Ziwen Huang}
\affiliation{Department of Physics and Astronomy, Northwestern University, Evanston, IL, 60208, USA}
\author{Jens Koch}
\affiliation{Department of Physics and Astronomy, Northwestern University, Evanston, IL, 60208, USA}
\author{Andrew A. Houck}\email{aahouck@princeton.edu}
\affiliation{Department of Electrical Engineering, Princeton University, Princeton, New Jersey 08540, USA}

\date{\today}

\begin{abstract}
We use the quasienergy structure that emerges when a fluxonium superconducting circuit is driven periodically to encode quantum information with dynamically induced flux-insensitive sweet spots. The framework of Floquet theory provides an intuitive description of these high-coherence working points located away from the half-flux symmetry point of the undriven qubit. This approach offers flexibility in choosing the flux bias point and the energy of the logical qubit states as shown in [\textit{Huang et al., 2020}]. We characterize the response of the system to noise in the modulation amplitude and DC flux bias, and experimentally demonstrate an optimal working point which is simultaneously insensitive against fluctuations in both. We observe a 40-fold enhancement of the qubit coherence times measured with Ramsey-type interferometry at the dynamical sweet spot compared with static operation at the same bias point.

\end{abstract}


\maketitle
\section{Introduction}

Superconducting circuits with a single degree of freedom, such as the transmon and fluxonium qubits, constitute the backbone of present solid-state quantum computers~\cite{martinis,kjaergaard,krantz,preskill,arute,havlicek}. The development of qubits equipped with a higher dimensional configuration space has recently opened a way to host protected states and represents a new platform to explore a wide range of fundamental quantum phenomena~\cite{bell,kou,kalashnikov,gyenis}. An alternative path to increase the effective dimensionality of a qubit state without introducing complex multinode circuits is to encode quantum information into time-dependent states.  When an artificial atom is irradiated by an intense field, the emerging system provides controllable states with desirable coupling to the environment~\cite{huang}. Such strongly driven superconducting circuits have been extensively studied in the context of Landau-Zener interference~\cite{ashhab,shevchenko,oliver,sillanpa,silveri}, sideband transitions~\cite{beaudoin,strand,li,naik},  Floquet-engineering~\cite{deng1,deng2,zhang}, on-demand dynamical control of operating points~\cite{didier,hong,fried,didier2}, tunable coupling schemes~\cite{bertet,niskanen,mckay,wu,reagor,mundada}, and many-body interaction in quantum simulators~\cite{roushan,wang,cai,kyriienko,sameti}. 

Here, we experimentally characterize a fluxonium qubit under strong external flux modulation and use the Floquet states to store quantum information.  One advantage of this approach is that we can create a dynamical flux-insensitive working point by harnessing the avoided crossings between quasienergy levels. This allows us to systematically realize favorable working bias points with \textit{in situ} tunable transition energies to help avoid frequency-crowding challenges in multiqubit processors.  Floquet theory provides a powerful tool to intuitively describe the behavior of this driven system\cite{floquet,grifoni,chu,son,silveri2}. 

The key idea of this formalism is that, when the evolution of a system is governed by a time-periodic Hamiltonian $H(t) = H(t + 2\pi/\Omega)$ with modulation frequency $\Omega$, there exists a special set of solutions  $|\psi_\alpha(t)\rangle$  of the time-dependent Schr\"odinger equation such that $|\psi_\alpha(t)\rangle$ can be expressed as the product of a dynamical phase factor and a time-periodic state $|\Phi_{\alpha}(t)\rangle$ in the form of $|\psi_\alpha(t)\rangle = e^{-i\epsilon_\alpha t/\hbar}|\Phi_{\alpha}(t)\rangle$, where $|\Phi_{\alpha}(t)\rangle = |\Phi_{\alpha}(t + 2\pi/\Omega)\rangle$. These states are the quasi-stationary Floquet states with $\alpha=0,1,\ldots$ labeling the solutions, while the characteristic energy $\epsilon_\alpha$ appearing in the dynamical phase factor is the quasienergy of the state. Importantly, $|\Phi_{\alpha}(t)\rangle$ has the same time-periodicity as the drive, and thus can be expressed as a discrete Fourier series in terms of the harmonics of the drive frequency $|\Phi_{\alpha}(t)\rangle=\sum_n e^{in\Omega t} |\phi_{\alpha}^{(n)}\rangle$. The unnormalized Fourier coefficients $|\phi_{\alpha}^{(n)} \rangle$ represent the atomic wavefunctions dressed by the periodic drive~\cite{wilson1,wilson2} and are called  quasi-wavefunctions.  In this work, we use the Floquet states $|\Phi_{\alpha}(t)\rangle$ as the computational basis states for our qubit.

The outline of the paper is as follows. In Sec.~\ref{The Floquet-fluxonium qubit} we describe the driven fluxonium qubit and the corresponding emerging Floquet states based on numerical diagonalization of the underlying Floquet Hamiltonian. In Sec.~\ref{Mapping the quasi-energy structure} we present the experimental signatures of Floquet states: Floquet polaritons and excitations of sidebands. In Sec.~\ref{Coherence of the Floquet states} we present time-domain measurements that demonstrate enhanced qubit coherence away from the high symmetry point resulting from a dynamical sweet spot. We provide additional data and theoretical details in the Appendices.

\section{The Floquet-fluxonium qubit}\label{The Floquet-fluxonium qubit}

In this work, we present a strongly-driven fluxonium qubit and show that coherence can be enhanced compared to static operation by operating at the dynamical sweet spots. Our qubit is operated in the light fluxonium regime~\cite{nguyen}, where a small Josephson junction is shunted by a relatively small capacitance and a large inductance. The qubit can be biased with an external flux that can be periodically-modulated. The driven fluxonium Hamiltonian is:
\begin{equation}\label{eq:hamiltonian}
    H(t) = 4E_Cn^2 - E_J\cos\varphi +\frac{1}{2}E_L\left(\varphi-2\pi\frac{\Phi_\mathrm{ext}(t)}{\Phi_0}\right)^2,
\end{equation}
where $\varphi$ and $n$ are the phase and charge operators, $E_J / h =$ 2.65 GHz  is the Josephson energy, $E_L/ h$ = 0.54 GHz is the inductive energy and $E_C / h = $ 1.17 GHz is the capacitive energy. Lastly, $\Phi_\mathrm{ext}(t)$ denotes the time-dependent magnetic flux threaded through the loop formed by the junction and inductor, and $\Phi_0$ is the magnetic flux quantum. The qubit is capacitively coupled to a tantalum-based~\cite{place} coplanar resonator with a resonance frequency of 7.30 GHz, allowing us to perform dispersive readout~\cite{blais}. 

DC flux bias control is commonly used to tune the transition energies of the qubit, but also adversely affects coherence by exposing the qubit to ubiquitous $1/f$ flux noise. To protect against this noise, it is necessary to set the qubit energy to an extremum value (called a first-order-insensitive \textit{static sweet spot}), which diminishes the advantage of the tunability of the device. Fortunately, by leveraging modulation techniques, we have the capability to recover -- under certain conditions -- the flexibility of choosing the optimal working point of the qubit and create an in-situ tunable \textit{dynamical sweet spot}.  

We consider the case when the magnetic flux is modulated with a single frequency $\Omega$ and amplitude $\xi$ around the static flux bias point $\Phi_\mathrm{ext}^0$, i.e., $\Phi_\mathrm{ext}(t) = \Phi_\mathrm{ext}^0 +\xi\cos(\Omega t)$. Floquet's theorem offers a natural description of such systems, where the Fourier components $|\phi_{\alpha}^{(n)} \rangle$ and the corresponding quasienergies $\epsilon_\alpha$ describe the dynamics. Importantly, as we illustrate below, these emerging Floquet states are suitable for quantum information processing in a manner similar to stationary qubits. To get an intuitive picture for the structure of the fluxonium Floquet states~\cite{rudner}, it is advantageous to introduce the time-averaged spectral function  $A_\alpha(\omega)=\sum_n\langle\phi_{\alpha}^{(n)}|\phi_{\alpha}^{(n)}\rangle\delta(\epsilon_\alpha+n\Omega-\omega)$, which captures the energy distribution of the spectral weight of a given Floquet state over multiple sidebands.

\begin{figure*}
    \centering
    \includegraphics[width = \textwidth]{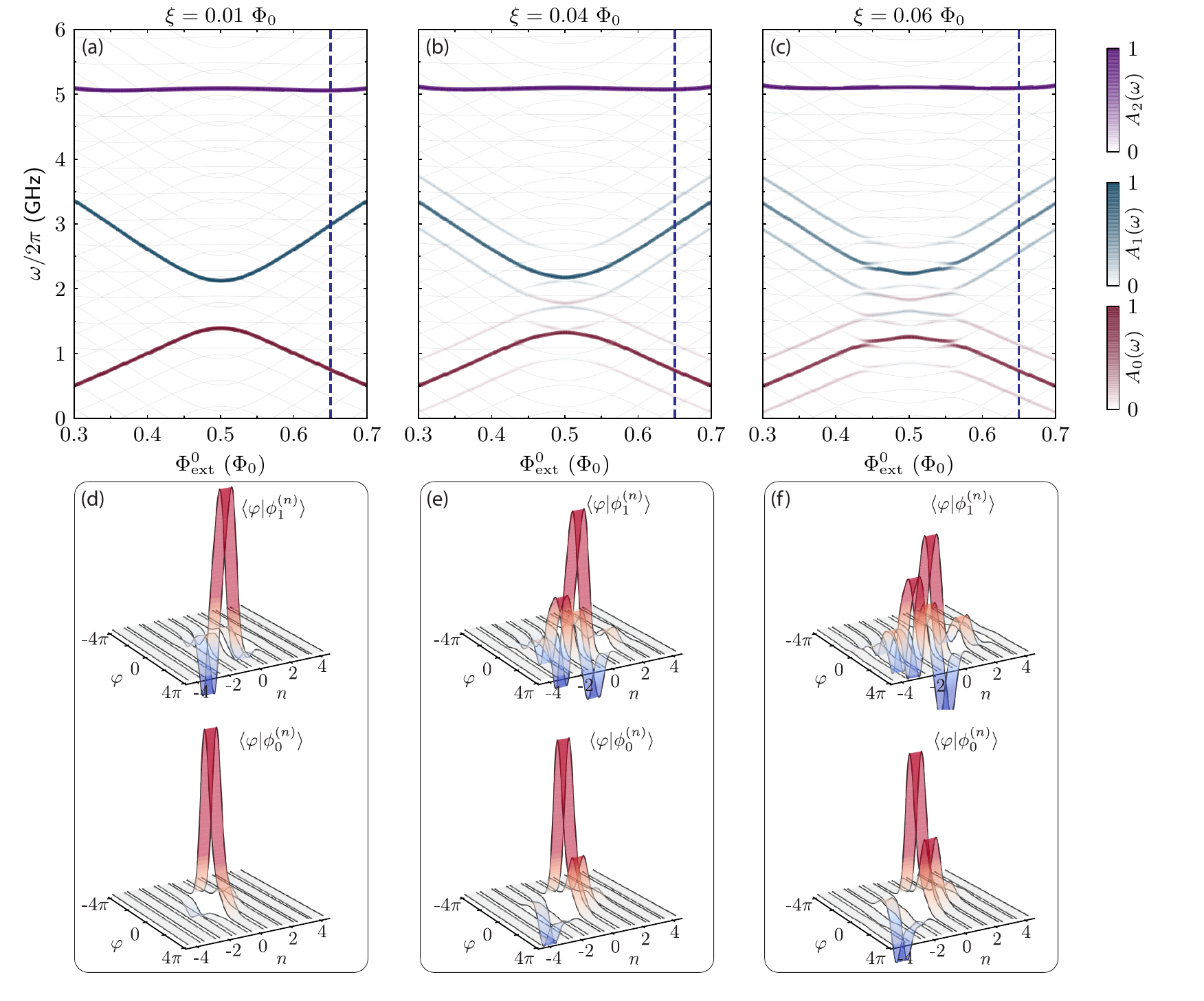}
    \caption{Numerically calculated Floquet quasienergy spectrum and quasi-wavefunctions. (a-c) Gray lines show the quasienergy $\epsilon_\alpha$ as function of external flux and flux drive amplitude $\xi$ at a fixed flux drive frequency of $\Omega/2\pi$ = 0.4 GHz. The weight of the time-averaged spectral functions $A_\alpha(\omega)$ are visualized by colored lines (red, blue, purple). For illustration purposes, we replaced the Dirac delta functions with Kronecker deltas in the definition of the spectral functions: $A_\alpha(\omega)=\sum_n\langle\phi_{\alpha}^{(n)}|\phi_{\alpha}^{(n)}\rangle\delta_{\epsilon_\alpha+n\Omega,\omega}$.  The distribution of the spectral weight among different sidebands depends on the flux dispersion of the states and strength of the flux modulation. (d-f) The Fourier components  of the Floquet states $\langle\varphi|\phi_{\alpha}^{(n)}\rangle$ visualized on a two-dimensional space spanned by the canonical phase variable and the sideband index. The spread of the wavefunctions into sidebands increases with the modulation strength.}
    \label{fig:2D states}
\end{figure*}

In \autoref{fig:2D states}, we present the quasienergies, quasi-wavefunctions and time-averaged spectral weights of the Floquet states as a function of DC flux bias and AC modulation amplitude as obtained numerically by diagonalizing the time-independent Floquet Hamiltonian. First, the quasienergies of the driven fluxonium (gray lines in \autoref{fig:2D states}a-c), can be characterized by an infinite set of multiphoton resonances, with period corresponding to the drive frequency. This redundancy of the quasienergies is the result of the discrete time-translation invariance of the driven-qubit Hamiltonian. 

To illustrate how the drive strength affects the Floquet states, we consider the time-averaged spectral function (colored lines in \autoref{fig:2D states}a-c), as well as the Fourier components of the ground and first excited Floquet states at different driving amplitudes (\autoref{fig:2D states}d-f). In the weak modulation-strength limit, the Floquet states resemble the static fluxonium states with spectral weight primarily located in a single harmonic and energy levels equivalent to the bare fluxonium qubit (\autoref{fig:2D states}a). Consequently, the wavefunctions are localized in a single mode ($n=0$), and they have the same shape in configuration space as the original fluxonium wavefunctions (\autoref{fig:2D states}d). We observe that as the drive power is increased, the spectral weight of the Floquet states spreads into sidebands (\autoref{fig:2D states}b,e). Intriguingly, at even higher powers, the Floquet states only have  a weak resemblance to the original fluxonium states, and the quasienergies form a complex pattern with spectral weight widely spread over numerous sidebands (\autoref{fig:2D states}c,f). 

Finally, the spectrum exhibits avoided crossings at various flux-bias values due to hybridization between Floquet sidebands. At these special parameter values~\cite{huang}, the system has dynamical sweet spots with transition energies first-order insensitive to the DC flux bias. In this work, we focus on these regions, where the Floquet-fluxonium qubit can be operated while maintaining high coherence.

\section{Measuring the quasi-energy spectrum}\label{Mapping the quasi-energy structure}
\begin{figure}
    \centering
    \includegraphics[width = \columnwidth]{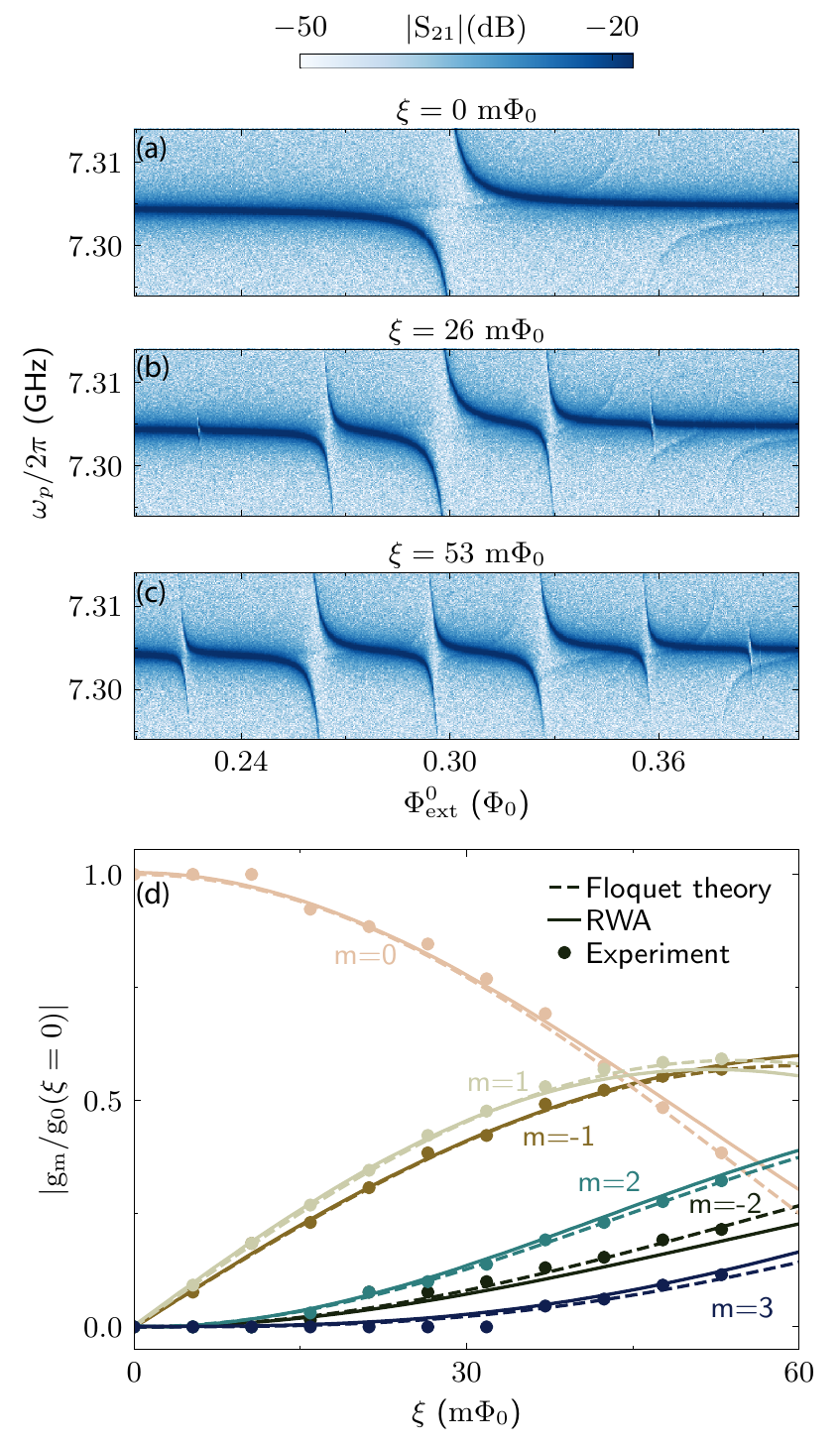}
    \caption{Floquet polariton states as a function of flux drive amplitude. (a-c) Vacuum Rabi splitting of the resonator with the Floquet sidebands measured in the transmission spectrum when $\Omega/2\pi$ = 0.2 GHz. (d) Experimentally extracted normalized coupling rates $g_m$ (solid dots) for the various sidebands $m$ with calculations based on a rotating-wave-approximation model (solid lines) and Floquet theory (dotted-dashed lines). Generally, as the modulation amplitude is increased, the spectral weight shifts towards higher-order sidebands.}
    \label{fig:polariton}
\end{figure}
In order to coherently control this strongly-driven Floquet qubit, we must first experimentally characterize and verify the quasienergy spectrum. We first focus on parametrically induced vacuum Rabi oscillations between a single mode of the readout cavity and the Floquet states. Similar to standard transmission measurements in the strong-coupling regime of superconducting qubits~\cite{wallraff}, we measure the response of the cavity as a function of DC flux bias. When one of the quasienergy differences is in resonance with the cavity frequency, the coherent exchange of a photon between the qubit and the cavity is indicated by avoided crossings in the transmission signal, and the coupling rate $g$ is captured by the size of the crossing. We systematically characterize these Floquet polariton states~\cite{clark} as a function of modulation amplitude, while keeping the drive frequency constant. As \autoref{fig:polariton}a shows, the transmission data features a single avoided crossing in the absence of flux modulation corresponding to the transition from the ground state to the third excited state of the fluxonium qubit. When the amplitude of the drive is increased (\autoref{fig:polariton}b,c), the spectral weight splits into higher-order sidebands, enhancing the dipole coupling between the bands detuned by multiples of the flux modulation. This is indicated by the emergence of additional avoided crossings of  the cavity with the higher-order sidebands as a function of the modulation amplitude. As the magnitude of the dipole matrix element is proportional to the amplitude of the wavefunctions, measuring the strength of the avoided crossing enables us to directly characterize the redistribution of the wavefunction in the different sidebands. The measured coupling rates (\autoref{fig:polariton}d) reveal that the spectral weight continuously transfers to the higher order sidebands as the drive strength is increased.

\begin{figure*}
    \centering
    \includegraphics[width = \textwidth]{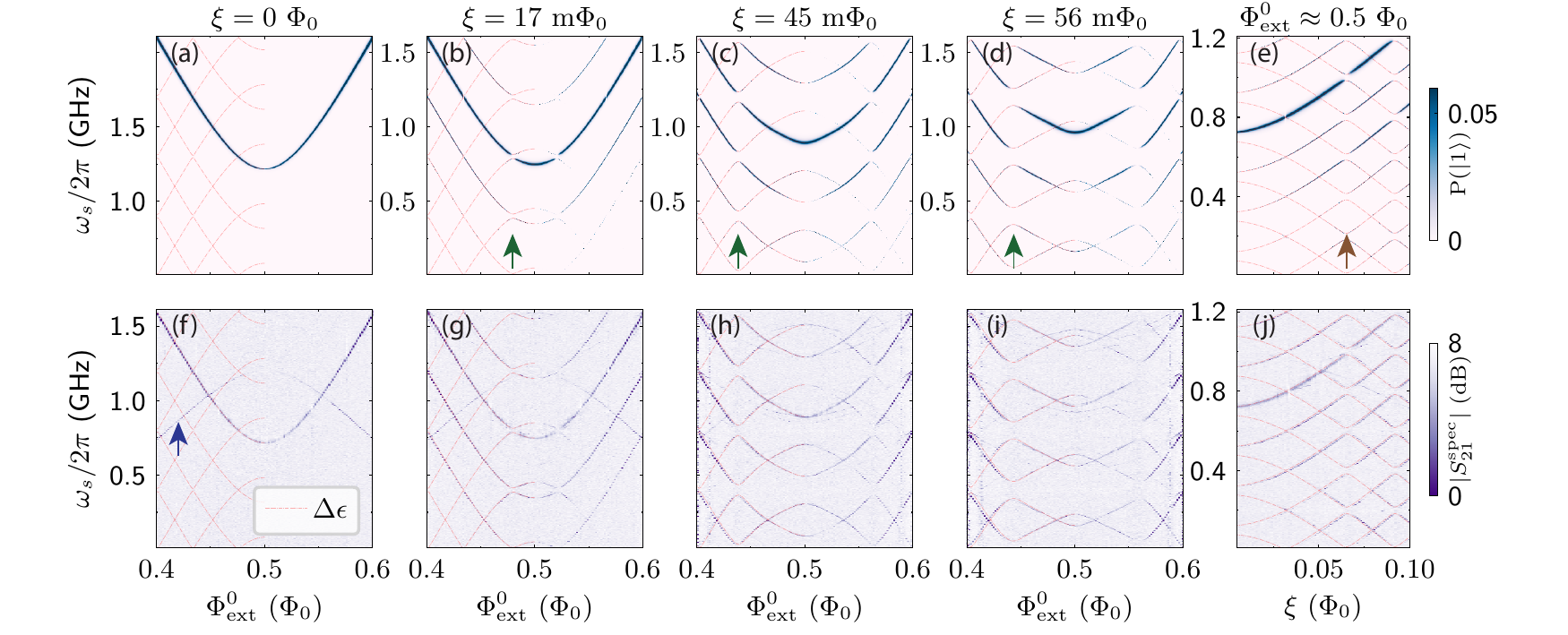}
    \caption{Spectroscopy measurements of the Floquet states. (a-e) Calculated dispersion of the quasienergies (red lines), and the simulated  qubit excitation during spectroscopy show dynamical sweet spots away from half flux bias. The steady-state simulation utilizes the Floquet master equation and is used to mimic the spectroscopy signals shown in the lower panels. Both green and brown arrows indicate dynamical sweet spots. The spots with first-order insensitivity to the flux bias are shown by green arrows, while the brown arrow indicates insensitivity to modulation amplitude. A double sweet spot must be simultaneously insensitive to both the dephasing channels. (f-j) Two-tone spectroscopy data on the driven fluxonium in the vicinity of half flux quantum. The measured transition energies match with the calculated quasienergy differences (red dashed lines) and are well reproduced in the steady state simulation. With increasing drive amplitude, more transitions are observed in the data due to the splitting of spectral weight, activating sideband transitions. The blue arrow marks the multi-photon transition between the cavity and higher qubit levels}
    \label{fig:spectroscopy}
\end{figure*}
We now proceed to spectroscopic measurements to map out the dynamical sweet spots in the driven fluxonium qubit, which are first-order insensitive against fluctuations of the DC flux bias and AC flux modulation amplitude. Here, we focus on the flux region close to half flux quantum, and perform two-tone spectroscopy in the low-energy region by monitoring the cavity transmission while an additional weak tone is applied to the flux-modulated system. Due to the ac-Stark effect, occupation of the qubit's excited state shifts the cavity's transition frequency. This leads to a reduction in our transmission signal when the qubit is excited. For comparison, we use the Floquet master equation to compute the steady-state qubit population during the spectroscopy experiment (\autoref{fig:spectroscopy}a-e) and which agrees well with the spectroscopy data observed in \autoref{fig:spectroscopy}f-j. In the undriven case (\autoref{fig:spectroscopy}a,f), the spectral weight in the sidebands is absent, and thus, the spectroscopic data shows a single transition with a static flux sweet spot at a half-flux quantum. The additional transition observed in the experiment (blue arrow in \autoref{fig:spectroscopy}f) can be accounted for as a multi-photon transition between the cavity and higher qubit level. Similar to the previously discussed transmission measurements, as the amplitude of the flux modulation is increased, the spectral weight propagates into the harmonics of the drive frequency. This enables transitions between the sidebands of the Floquet states due to the weak probe field. The obtained low-energy spectra  (\autoref{fig:spectroscopy}g-i) demonstrate the growing number of allowed transitions between these harmonics as predicted by the Floquet theory. This behavior is even more apparent in \autoref{fig:spectroscopy}j, which shows the excitation spectrum as a function of drive amplitude at a fixed DC flux bias.

Importantly, the flux modulation not only redistributes the spectral weight of the qubit state into sidebands but also changes the flux dispersion of the quasienergies. This can be understood as an interaction between the sidebands, which exhibit avoided crossings with an energy splitting proportional to the flux drive amplitude. Such avoided crossings create dynamical sweet spots, where the derivative of the quasienergy differences vanishes, offering first-order insensitivity against DC flux noise at tunable flux bias values (green arrows in \autoref{fig:spectroscopy}). 

We emphasize that by coupling the qubit to the flux drive, we also introduce additional coupling to fluctuating parameters, for instance, drive frequency, amplitude and phase, which can lead to dephasing. Given the frequency stability of commercial microwave generators, we focus on the dephasing caused by fluctuations in the drive amplitude. The DC-flux value corresponding to the dynamical sweet spot is strongly dependent on the amplitude of the drive $\xi$ (green arrows in \autoref{fig:spectroscopy}), which enables noise in the drive amplitude to potentially degrade the flux insensitivity of the dynamical sweet spots -- unless those are also insensitive to drive amplitude fluctuations. In other words, first-order insensitivity against noise in the DC flux bias is necessary but not sufficient to preserve the phase of the qubit. An example of a sweet spot for the amplitude of the drive is shown in \autoref{fig:spectroscopy}e and j (brown arrow). A double sweet spot, which has vanishing derivatives with respect to both DC flux bias and AC drive amplitude, provides simultaneous insensitivity to the DC bias fluctuations and the AC flux noise~\cite{didier,didier2,huang}. Fortunately, as shown below, such double sweet spots under flux modulation can be found in the drive parameters. 

\section{Coherence of the Floquet states}\label{Coherence of the Floquet states}
\begin{figure*}
    \centering
    \includegraphics[width = \textwidth]{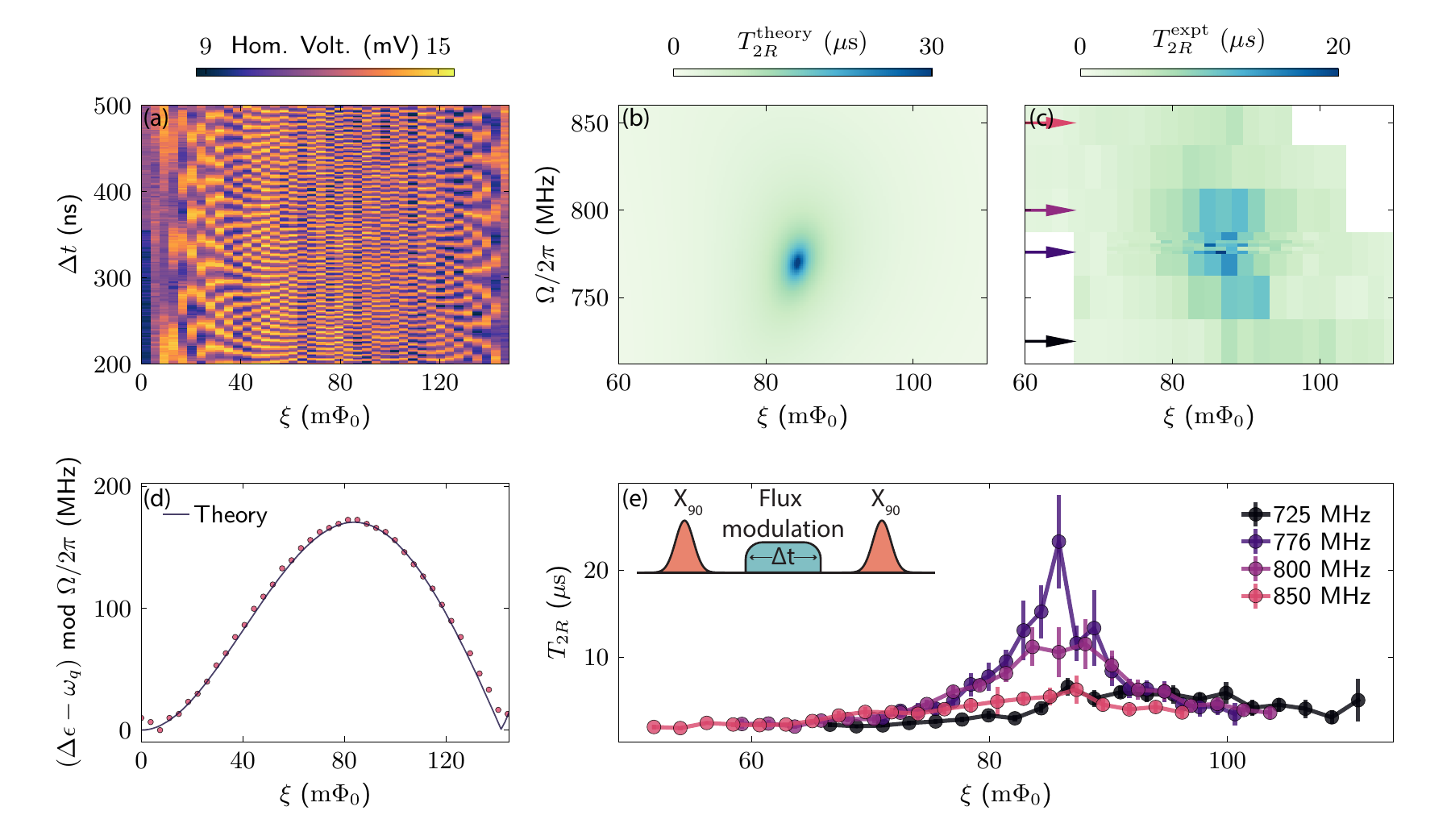}
    \caption{Floquet states in the time-domain. (a) The measured homodyne voltage signal at the end of the Ramsey-type protocol (pulse scheme is depicted in panel e inset). The homodyne signal is proportional to the ground state population of the static qubit, and the oscillations are the result of the dynamical phase accumulation due to the time evolution of the Floquet states. (d) The extracted rate of phase accumulation matches perfectly with the numerical results obtained from Floquet theory. (b) Calculation~\cite{huang} of the pure dephasing rate for the Floquet states shows a dynamical sweet spot in the $(\xi,\Omega)$ flux-drive-parameter space. (c) Experimental measurements of Ramsey dephasing times $T_{\mathrm{2R}}$ for different modulation strengths and frequencies demonstrates similar behavior. At the simultaneous sweet spot a 40 times enhancement of the coherence time is observed compared to the undriven case. (e) Measured $T_{\mathrm{2R}}$ as a function of the drive amplitude at various drive frequencies (corresponding to the linecuts in panel c). All time-resolved experimental data and calculations are performed at $\Phi_\mathrm{ext}^0 = 0.451$~$\Phi_0$.}
    \label{fig:coherence}
\end{figure*}
With the location of sweet spots established, we can now present the central result of this paper: time-domain measurements of the coherence properties of the Floquet states.  We measure the acquired dynamical phase of the Floquet states in a Ramsey-type protocol. As \autoref{fig:coherence}e inset shows, we initially prepare an equal superposition of the ground and first excited state of the undriven qubit by applying an $X_{90}$ gate to the ground state. We then adiabatically turn on the flux modulation such that the system follows the instantaneous Floquet states \cite{deng1,deng2}, which creates an equal superposition of the ground and first excited Floquet states. Following this, the system evolves under a modulation with constant amplitude $\xi$ and frequency $\Omega$ for time $\Delta t$. At the end, the modulation is again adiabatically turned off, and the excited state population is measured after another $X_{90}$ gate. The measured time evolution of the qubit population for different driving amplitudes and the rates of phase accumulation are plotted in \autoref{fig:coherence}a,d. The frequency components of the oscillation are expected to follow the quasienergy difference $\Delta\epsilon+n\Omega$ displaced by the bare qubit frequency $\omega_0$ (due to the $X_{90}$ gates). The extremum of the quasienergy difference, i.e. the dynamical sweet spot can be found by comparing the measured frequency components to numerical simulation and the spectroscopic data.  

The presented Ramsey-type protocol can be used to determine the coherence of the Floquet states. By changing the length of the modulation pulse for extended periods and measuring the amplitude of the decaying oscillation, we can probe the driven qubit coherence. To find the double sweet spot, we measure the coherence both as function of drive amplitude and frequency. The time-domain measurements (\autoref{fig:coherence}c) reveal the presence of such a double sweet spot as predicted by the Floquet theory (in \autoref{fig:coherence}b). The double sweet spot provides a 40-fold enhancement of $T_{2R}$ at the cost of a 3.5-fold reduction in $T_1$ (refer \autoref{table1}). This result clearly demonstrates the potential of Floquet engineering for achieving ideal trade-offs between depolarization and dephasing in quantum processors.

\begin{table}[h]
\caption{Measured coherence times of the system. At the double sweet spot, Floquet drive decreases the $T_1$ of the fluxonium away from half flux bias by 3.5 times while increasing the $T_{2R}$ by 40 times.}
\begin{center}
\begin{ruledtabular}
\begin{tabular}{lll}
$\Phi_{\mathrm{ext}}^0$ $(\Phi_0)$ & $T_1$ ($\mu$s) & $T_{2R}$ ($\mu$s) \\\hline
$0.500$ (undriven) & $162 \pm 15$ & $76 \pm 5$\\ 
$0.451$ (undriven) & $91 \pm 10$ & $0.63\pm 0.07$ \\ 
$0.451$ (driven)  & $26 \pm 5$ & $23 \pm 5$  \\ 
\end{tabular}

\end{ruledtabular}
\end{center}
\label{table1}
\end{table} 

\section{Conclusions}
In this work, we have presented the steady-state response and time-resolved behavior of a fluxonium qubit under strong flux modulation. The measured spectroscopic features are in excellent agreement with numerical calculations based on Floquet theory, and clearly demonstrate the emergence of tunable dynamical sweet spots that can be used to preserve coherence away from the static sweet spot.  In particular, we engineer a dynamical sweet spot which is simultaneously first-order-insensitive against fluctuations in DC flux bias and AC modulation amplitude.  At this bias point, the coherence time approaches the measured value at the static sweet spot and is forty times greater than the coherence observed in static operation at the same bias point, away from the static sweet spot. Our findings open new possibilities to realize and control versatile superconducting circuits that combine the coherence benefits of operation at a sweet spot, while maintaining a degree of tunability.  

\section*{Acknowledgements}
This work is supported by Army Research Office Grant No.\ W911NF-19-1-0016. Devices were fabricated in the Princeton University Quantum Device Nanofabrication Laboratory and in the Princeton Institute for the Science and Technology of Materials (PRISM) cleanroom. The authors acknowledge the use of Princeton's Imaging and Analysis Center, which is partially supported by the Princeton Center for Complex Materials, a National Science Foundation (NSF)-MRSEC program (DMR-1420541).

\appendix

\section{Fluxonium spectrum}\label{App:FluxSpec}
\begin{figure}
    \centering
    \includegraphics[width = \columnwidth]{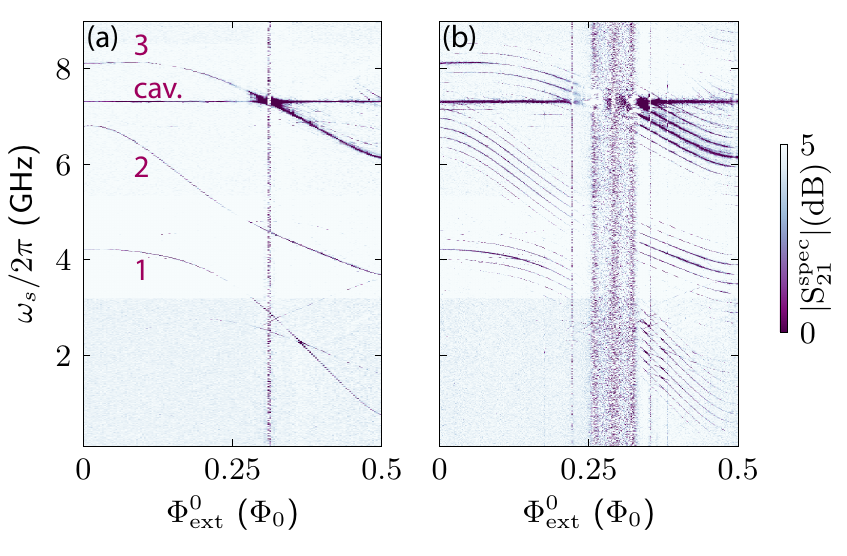}
    \caption{Two-tone spectroscopy data on the (a) un-driven (b) driven fluxonium w.r.t. external flux bias. The data show clear signatures of first three excited levels and the readout cavity. In the driven case, more transitions are observed in the data due to the splitting of spectral weight that activates sideband transitions (Drive frequency $\Omega/2\pi$ = 0.2 GHz).}
    \label{fig:FullSpec}
\end{figure}

\autoref{fig:FullSpec}a shows the two-tone spectroscopy data for the undriven fluxonium qubit as a function of DC flux bias. We extract the device parameters -- $E_L$, $E_J$ and $E_C$ -- based on the observed energy dispersion of the transitions. \autoref{fig:FullSpec}b presents transitions in the driven Fluxonium system. Qualitatively, this spectrum comprises of multiple copies of the spectrum in \autoref{fig:FullSpec}a shifted by the flux modulation frequency ($\Omega/2/\pi$ = 0.2 GHz). Furthermore, since we weakly probe the transmission of the cavity at it's frequency corresponding to the undriven qubit in the ground state, the vertical lines observed around $\Phi_{\mathrm{ext}}^0=0.3 \Phi_0$ correspond to the hybridization of the cavity which shifts the resonance frequency of the cavity. The driven fluxonium clearly exhibits multiple such vertical lines arising due to the hybridization of the cavity with Floquet sidebands.

\section{Extracting the dipole coupling of Floquet mode}\label{App:FloPol}
\subsection{Full Floquet theory}
We numerically solve the time-independent Floquet Hamiltonian for a fluxonium approximated with its 5 lowest energy states. This model follows the treatment provided in Ref.~\cite{son}  and accounts for multi-photon effects. To model the experiment presented in \autoref{fig:polariton}, we consider the dipole coupling between levels 0 and 3. The fluxonium qubit is capacitively coupled to the resonator. The coupling is described by a coupling term $g^\mathrm{cap} \hat{n}(a+a^\dagger)$, where $\hat{n}$ is the charge operator of the fluxonium, and $a (a^{\dagger})$ is the annihilation (creation) operator for the cavity. The coupling strength of the $m^{th}$ sideband transition is
\begin{equation}
g_{03}^m=g^\mathrm{cap}\langle\phi_3^{(m)}|\hat{n}|\phi_0^{(0)}\rangle.
\end{equation}

\subsection{Rotating Wave Approximation}
Considering a two-level system comprising of level 0 and level 3 of the fluxonium qubit and use the treatment provided in Ref.~\cite{silveri2}. The Hamiltonian can be written as 

\begin{equation}
H = \left[\omega_3+\zeta(\delta\Phi_\mathrm{ext}(t))\right]\sigma_z + [g + g'\cos(\Omega t)]\cos(\omega_p t)\sigma_x
\end{equation}
where $\omega_3(\Phi_{\mathrm{ext}}^0)$ and $g(\Phi_{\mathrm{ext}}^0)$ are the qubit transition energy and cavity-qubit coupling respectively for a DC flux bias of $\Phi_{\mathrm{ext}}^0$. The periodical flux modulation $\delta\Phi_\mathrm{ext}=\xi \cos(\Omega t)$ leads to a modulation in the qubit frequency and its coupling to the cavity. The qubit frequency modulation is distorted due to a non-linear energy-flux dispersion of the qubit $\zeta(\delta\Phi_\mathrm{ext})$. The modulation of the coupling can be linearly approximated with $g'\cos(\Omega t)$.

We transform the Hamiltonian into the interaction picture by using the unitary transformation of $U=e^{-i[\omega_3t + \eta(t)]\sigma_z}$, where $\eta(t)=\int_0^t\zeta(\delta\Phi_\mathrm{ext}(\tau))d\tau$. The Hamiltonian of the probed system becomes
\begin{align}
H^{(I)} &= [g + g'\cos(\Omega t)]\cos(\omega_p t)\nonumber \\ &\times[A(t)e^{i\omega_3t}\sigma_+ + A^*(t)e^{-i\omega_3t}\sigma_-],
\end{align}
where the dynamical phase factor is $A(t)=e^{i\eta(t)}$. After expressing the periodic phase factor with its Fourier components $A(t)=\sum_n A_n e^{in\Omega t}$, the Hamiltonian takes the form 
\begin{align}
H^{(I)} &= [g + g'\cos(\Omega t)]\cos(\omega_p t)\nonumber \\ &\times \sum_n [A_n e^{i(\omega_3+n\Omega)}\sigma_+ + \mathrm{h.c.}].
\end{align}

In the close proximity of multi-photon resonances, i.e $\omega_p\approx\omega_3+n\Omega$, we use the RWA approximation by neglecting fast-rotating driving terms:
\begin{align}
H^{(I)}_\mathrm{RWA} &=\left[\frac{g}{2}A_n + \frac{g'}{4}(A_{n-1} + A_{n+1})\right] \nonumber \\ &\times \left(e^{i(\omega_3+n\Omega-\omega_p)t}\right)\sigma_+ + \mathrm{h.c.}.
\end{align}
In the Schrodinger picture, this Hamiltonian reads as 
\begin{align}
H_\mathrm{RWA} &= [\omega_3+n\Omega-\omega_p]\sigma_z \nonumber \\&+ \left(\left[\frac{g}{2}A_n + \frac{g'}{4}(A_{n-1} + A_{n+1})\right] \sigma_+ + \mathrm{h.c.}\right),
\end{align}
which shows that the cavity-qubit coupling rate is modulated by the Fourier coefficients $A_n$, $A_{n-1}$ and $A_{n+1}$ where $A_n = \frac{\Omega}{2\pi}\int_0^{2\pi/\Omega}e^{-in\Omega t}e^{i\eta(t)}dt$.
The effective coupling of a Floquet mode with the cavity is thus given by 
\begin{equation}
g_n^\mathrm{RWA} = \left[\frac{g}{2}A_n + \frac{g'}{4}(A_{n-1} + A_{n+1})\right].
\end{equation}

\subsection{Experiment}
To extract the coupling strengths of Floquet polaritons (\autoref{fig:polariton}d), for each value of modulation strength ($\xi$), we fit the transmission data with the eigenenergies of the following manifold containing a single excitation in the cavity or the sidebands of $\Phi_3$ - 
\begin{align}
\label{eq:H_polariton}
H_\mathrm{1 exc} = \omega_c\left|c\right>\left<c\right| + \sum_{m=-2}^3 &(\omega_3+m\Omega+\delta_m)\left|m\right>\left<m\right| \nonumber \\ &+ g_{03}^{(m)} (\left|c\right>\left<m\right| + \left|m\right>\left<c\right|),
\end{align}
where $g_{03}^{(m)}$ and $\delta_m$ are the dipole coupling strength and AC stark shift of the corresponding sidebands. $\omega_c$ and $\omega_3$ are the resonance frequencies of the cavity and the third excited level respectively. $\left|c\right>$ corresponds to state with one excitation in the resonator and $\left|m\right>$ represent the normalized sidebands of $\Phi_3$. The flux dependence of $\omega_3$ is calibrated from the spectroscopy data shown in \autoref{fig:FullSpec}a. The model only has $g_{03}^{(m)}$ and $\delta_m$ as the fit parameters. 

\section{Extraction of Floquet $T_{2R}$}\label{App:T2R}
\begin{figure*}
    \centering
    \includegraphics[width = \textwidth]{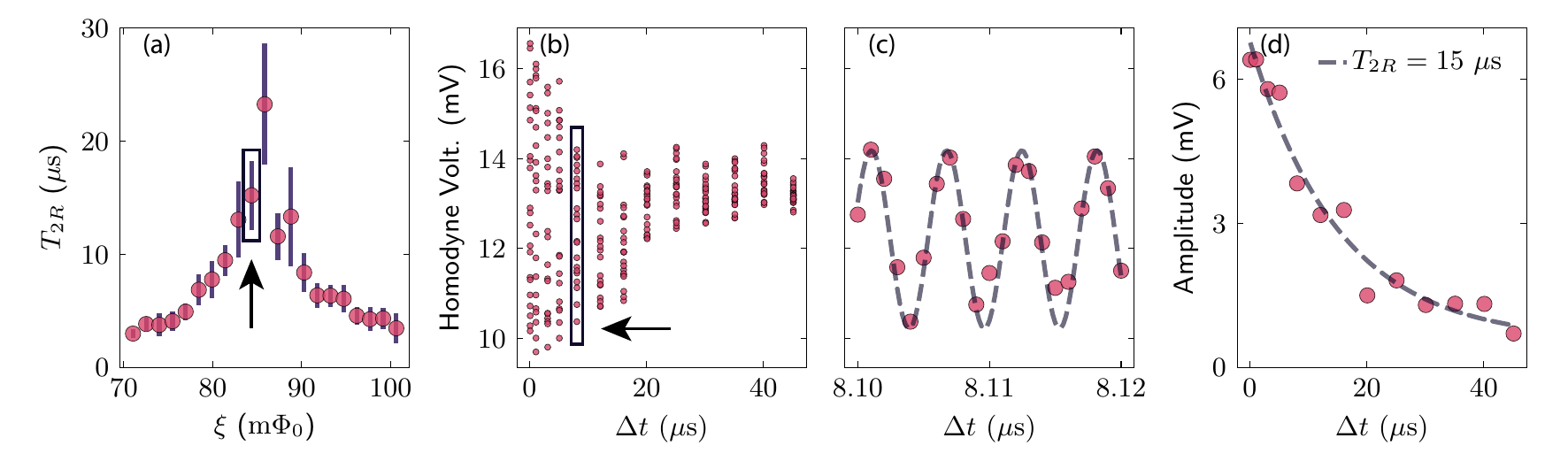}
    \caption{(a) Floquet $T_{2R}$ for $\Omega/2\pi$ = 776 MHz (also shown in \autoref{fig:coherence}e). (b) Time trace of the homodyne measurement corresponding to the boxed data point in (a). In order to maintain adequate sampling frequency to capture the accumulation of the dynamical phase of the Floquet state, we choose multiple windows with 20ns duration of delay times with varying delay offsets. (c) We show a zoomed-in data for the window boxed in (b). The amplitude of the oscillation does not change significantly within a single dense window of the delay times. (d) We fit the amplitudes of the oscillation at different delay offsets with an exponential function to obtain the $T_{2R}$ value.}
    \label{fig:T2Rextract}
\end{figure*}

We illustrate the extraction of $T_{2R}$ under flux modulation in \autoref{fig:T2Rextract}. The rate of accumulation of dynamical phase of the Floquet state in the vicinity of dynamical sweet spot is around 170 MHz (see \autoref{fig:coherence}d,e). To capture these fast oscillations, we vary time delay between the two $X_{90}$ pulses by 1 ns in a window of 20 ns. We further run the experiment for multiple such 20 ns windows offset by additional delay to precisely obtain the decay envelope (\autoref{fig:T2Rextract}b). The amplitude of the oscillation in each window (representative data in \autoref{fig:T2Rextract}c) can be fit with an exponential decay to obtain $T_{2R}$ as shown in \autoref{fig:T2Rextract}d.

\section{Dynamical sweet spots}\label{App:DynSweet}
\begin{figure*}
    \centering
    \includegraphics[width = \textwidth]{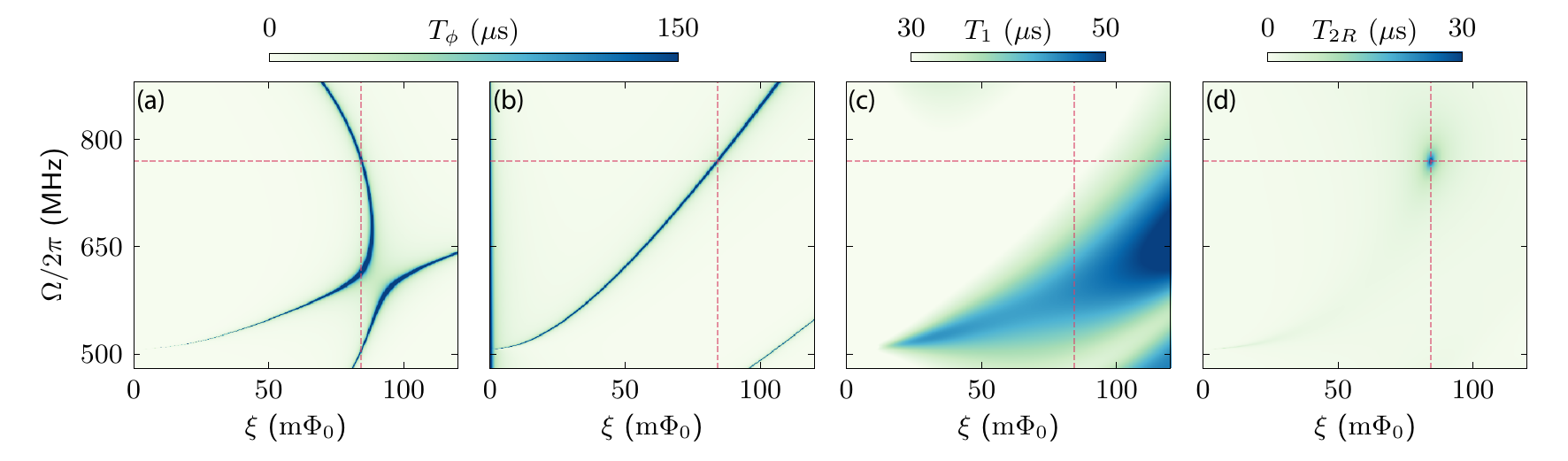}
    \caption{Calculation of the coherence times of the Floquet states show dynamical sweet regions in the $(\xi,\Omega)$ flux-drive-parameter space. Pure-dephasing time limited by 1/$f$ fluctuations in the (a) DC flux bias with $A_\mathrm{dc} = 7.5 \mu\Phi_0$, (b) modulation strength with $A_\mathrm{ac} = 6 \mu\Phi_0$. Coherence times (c) $T_1$ with $\mathrm{tan}\delta_c = 2.8\times10^{-6}$, $\mathrm{T} = 85 \mathrm{mK}$ and (d) $T_{2R}$. We achieve highest coherence enhancement for simultaneous insensitivity to the fluctuations in DC flux bias and modulation amplitude (intersection of dashed lines).}
    \label{fig:theoryTphi}
\end{figure*}

Using the formalism described in Ref.~\cite{didier2, huang}, we theoretically show the dependence of pure-dephasing and relaxation times on the modulation frequency and amplitude in \autoref{fig:theoryTphi}.

First, it is well established that the DC flux randomly fluctuates over time, which is associated with a 1/$f$ noise spectrum ~\cite{ithier,nguyen}. Second, early explorations of qubit dephasing under drives ~\cite{didier,didier2,hong,reagor} implies that ac flux noise also has a low-frequency nature. We assume that the fluctuation in ac modulation amplitude $\xi$, also has a 1/$f$ spectrum. Specifically, we assume the following noise spectra
\begin{align}
S_\mathrm{dc}(\omega)=&\,\int_{-\infty}^{\infty}\mathrm{d}t\, e^{i\omega t}\langle \delta \Phi_\mathrm{ext}^0(t)\delta\Phi_\mathrm{ext}^0(0)\rangle = 2\pi A^2_\mathrm{dc}/|\omega|,\nonumber\\
S_\mathrm{ac}(\omega)=&\,\int_{-\infty}^{\infty}\mathrm{d}t\, e^{i\omega t}\langle \delta \xi(t)\delta\xi(0)\rangle = 2\pi A^2_\mathrm{ac}/|\omega|.
\end{align}
Here, $A_\mathrm{dc}$ and $A_\mathrm{ac}$ are used to denote the strengths of the low-frequency noise. 

We further assume that the high-frequency noise mainly originates from the dielectric loss that causes qubit depolarization. The dielectric loss is coupled to the qubit by the fluxonium's phase operator $\varphi$ ~\cite{nguyen}. The associated noise spectrum is assumed to be
\begin{align}
S_\mathrm{diel} (\omega) &= \int_{-\infty}^{\infty}\mathrm{d}t\, e^{i\omega t} \langle \hat{O}_\mathrm{diel}(t)\hat{O}_\mathrm{diel}(0)\rangle \nonumber\\  &= \frac{\hbar \omega^2\mathrm{tan}\delta_c}{8E_C}\left[\coth\left(\frac{\hbar\omega}{2k_\mathrm{B}T}\right)+1\right],
\end{align}
where $\hat{O}_\mathrm{diel}$ is the noise operator for the dielectric loss. 

To derive the rates of decoherence of the driven qubit induced by these noise sources, we employ a Bloch-Redfield master equation, similar to  the treatment performed in Ref.~\cite{huang}. The interaction between the qubit and the noise sources is described by
\begin{align}
H_\mathrm{int} = \varphi \hat{O}_\mathrm{diel} + E_L\varphi \,\frac{2\pi\delta\Phi^0_\mathrm{ext} }{\Phi_0} + E_L\varphi\cos(\Omega t) \,\frac{2\pi\delta\xi}{\Phi_0}.
\end{align}
Note that all three terms in the equation above contain the operator $\varphi$. The decoherence rates are closely related to matrix elements of $\varphi$ in the basis of Floquet states $\vert \Phi_\alpha(t)\rangle$ ($\alpha=0,1$), i.e., $\varphi_{\alpha\alpha'}(t) = \langle \Phi_\alpha(t)\vert \varphi\vert \Phi_{\alpha'}(t)\rangle$, or more importantly, their Fourier coefficients 

\begin{align}
\varphi_{\alpha\alpha'}^{(k)} = \frac{\Omega}{2\pi}\int_0^{2\pi/\Omega} \mathrm{d}t \,e^{ik\Omega t} \varphi_{\alpha\alpha'}(t).
\end{align}

The depolarization and pure-dephasing rates are then given by

\begin{align}
\gamma_\pm &= \sum_{k\in\mathbb{Z}} |\varphi_{01}^{(k)}|^2 [S_\mathrm{diel}(k\Omega \mp \epsilon_{01}) + E^2_L \tilde{S}_\mathrm{dc}(k\Omega \mp \epsilon_{01})] \nonumber\\
&+ \frac{1}{4}\sum_{k\in\mathbb{Z}} (|\varphi_{01}^{(k+1)}+\varphi_{01}^{(k-1)}|^2)E_L^2 \tilde{S}_\mathrm{ac}(k\Omega\mp \epsilon_{01}),
\end{align}

\begin{widetext}
\begin{align}
\label{gamma_phi}
\gamma_\phi &= \sqrt{|\ln \omega_\mathrm{ir}t_\mathrm{m}|} \sqrt{E_L^2 \left(\frac{2\pi A_\mathrm{dc}}{\Phi_0}\right)^2|\varphi_{11}^{(0)}-\varphi_{00}^{(0)}|^2+\frac{1}{4}E_L^2\left(\frac{2\pi A_\mathrm{ac}}{\Phi_0}\right)^2|\varphi_{00}^{(1)}+\varphi_{00}^{(-1)}-\varphi_{11}^{(1)}-\varphi_{11}^{(-1)}|^2}\nonumber\\
&+\frac{1}{2}\sum_{k\neq 0} |\varphi_{11}^{(k)}-\varphi_{00}^{(k)}|^2 [S_\mathrm{diel}(k\Omega) + E^2_L \tilde{S}_\mathrm{dc}(k\Omega)] \nonumber\\
&+ \frac{1}{8}\sum_{k\neq 0} (|\varphi_{00}^{(k+1)}+\varphi_{00}^{(k-1)}-\varphi_{11}^{(k+1)}-\varphi_{11}^{(k-1)}|^2)E_L^2 \tilde{S}_\mathrm{ac}(k\Omega).
\end{align}
\end{widetext}
Above, we defined the reduced noise spectrum $\tilde{S}_\mathrm{dc}(\omega)=S_\mathrm{dc}(\omega)(2\pi/\Phi_0)^2$ and similarly for $\tilde{S}_\mathrm{ac}(\omega)$. We use $\omega_\mathrm{ir}$ to denote the infrared cutoff frequency for the noise, and $t_m$ for the characteristic measurement time.

Similar to the discussion in Ref.~\cite{huang}, we also show
\allowdisplaybreaks
\begin{align}
&\varphi_{11}^{(0)}- \varphi_{00}^{(0)} \sim \frac{\partial \epsilon_{01}}{\partial \Phi^0_\mathrm{ext}},\\
\label{strength_derivative}
&\varphi_{11}^{(1)}+\varphi_{11}^{(-1)}-\varphi_{00}^{(1)}-\varphi_{00}^{(-1)} \sim \frac{\partial \epsilon_{01}}{\partial \xi}.
\end{align}

\emph{Proof:} We invoke perturbation theory to prove the two relations shown above. For the first relation, we assume  that a perturbation term $\delta \Phi^0_\mathrm{ext}E_L\varphi$ is added to the driven qubit's Hamiltonian. It is straightforward to evaluate the first-order change in quasi-energy difference as

\begin{align}
\delta \epsilon_{01} =&\,  \left(\frac{2\pi}{\Phi_0}\right)\delta\Phi^0_\mathrm{ext}E_L \nonumber\\ &\times \frac{\Omega}{2\pi}\int_{0}^{2\pi/\Omega} dt\,\left[ \langle \Phi_1(t)\vert \varphi\vert \Phi_1(t)\rangle - \langle \Phi_0(t)\vert \varphi\vert \Phi_0(t)\rangle\right]\nonumber\\
=&\,\left( \frac{2\pi}{\Phi_0}\right)E_L\delta\Phi^0_\mathrm{ext} (\varphi_{11}^{(0)}-\varphi_{00}^{(0)}).
\end{align}

Therefore, we have $\partial \epsilon_{01}/\partial \Phi^0_\mathrm{ext} = ({2\pi}/{\Phi_0})E_L (\phi_{11}^{(0)}-\phi_{00}^{(0)})$. We can prove Eq.~\eqref{strength_derivative} similarly. The perturbation term is hereby chosen to be $\delta \xi\, E_L \varphi\cos(\Omega t)$, and the first-order correction is 

\begin{align}
\delta \epsilon_{01} =&\, \left(\frac{2\pi}{\Phi_0}\right)\delta\xi E_L \times \frac{\Omega}{2\pi}\int_{0}^{2\pi/\Omega} dt\,\cos(\Omega t)\nonumber\\ &\times\left[ \langle \Phi_1(t)\vert \varphi\vert \Phi_1(t)\rangle - \langle \Phi_0(t)\vert \varphi\vert \Phi_0(t)\rangle\right]\nonumber\\
=&\, \left(\frac{\pi}{\Phi_0}\right)\delta\xi E_L (\varphi_{11}^{(1)}+\varphi_{11}^{(-1)} - \varphi_{00}^{(1)}-\varphi_{00}^{(-1)}).
\end{align}

Then we derive the result $\partial \epsilon_{01}/\partial \xi = (\pi/\Phi_0)E_L (\varphi_{11}^{(1)}+\varphi_{11}^{(-1)} - \varphi_{00}^{(1)}-\varphi_{00}^{(-1)})$. Using these relations, we can rewrite Eq.~\eqref{gamma_phi} as
\begin{align}
\gamma_\phi = &\,  \sqrt{|\ln \omega_\mathrm{ir}t_\mathrm{m}|}\sqrt{A^2_\mathrm{dc}\left|\frac{\partial \epsilon_{01}}{\partial \Phi^0_\mathrm{ext}} \right|^2+A^2_\mathrm{ac} \left| \frac{\partial\epsilon_{01}}{\partial\xi}\right|^2}\nonumber\\
&+\frac{1}{2}\sum_{k\neq 0} |\varphi_{11}^{(k)}-\varphi_{00}^{(k)}|^2 [S_\mathrm{diel}(k\Omega) + E^2_L \tilde{S}_\mathrm{dc}(k\Omega)] \nonumber\\
&+ \frac{1}{8}\sum_{k\neq 0} (|\varphi_{00}^{(k+1)}+\varphi_{00}^{(k-1)}-\varphi_{11}^{(k+1)}-\varphi_{11}^{(k-1)}|^2) \nonumber \\ & \quad \quad \times E_L^2 \tilde{S}_\mathrm{ac}(k\Omega).
\end{align}
The double dynamical sweet spots correspond to the regimes in parameter space where both $\partial \epsilon_{01}/\partial \Phi^0_\mathrm{ext}$ and $\partial \epsilon_{01}/\partial \xi$ reach zero. We find that the observed reduction in $T_1$ for the driven qubit cannot be explained solely by the redistribution of the filter functions. However, this discrepancy can be reconciled by considering a bath temperature of $~85 \mathrm{mK}$. While prior experiments with a strong drive have observed similar heating of the device~\cite{zhang}, we note that further characterization is needed to understand the source of the heating in our devices which will be investigated in future studies.

\section{Redistribution of filter functions}\label{App:Filter}
The lowering of $T_1$ at the dynamical sweet spot compared with that of the undriven qubit is expected to be related to the increase of $\sum_k |\varphi_{01}^{(k)}|^2$, which are the filter-function weights indicating the sensitivity of the qubit to the dielectric loss. In fact, there exists a trade-off between the weights for pure-dephasing $\sum_k |\varphi_{11}^{(k)}-\varphi_{00}^{(k)}|^2/2$ and those for depolarization $\sum_k |\varphi_{01}^{(k)}|^2$ ~\cite{huang}. Therefore, when the pure-dephasing weights are suppressed at the sweet spot, leading to an increase in pure-dephasing time, the weights for depolarization increase. The proof of this conservation law as given in Ref.~\cite{huang} is provided below for completeness.

When the drive strength and frequency aren't sufficiently large enough to excite transitions to higher fluxonium states, we can conveniently describe the Floquet states in the basis of the lowest two eigenstates of the undriven qubit as $\vert \Phi_j(t)\rangle = \sum_{\sigma = 0,1} u_{j,\sigma}(t)\vert \sigma\rangle$, where $\vert \sigma\rangle$ denotes the eigenstate of the undriven qubit. 
\begin{align}
\varphi_{jj'}(t) &= \langle \Phi_j(t)\vert \varphi\vert \Phi_{j'}(t)\rangle \nonumber \\ &= \sum_{\sigma, \sigma'} u^*_{j,\sigma}(t) \bar{\varphi}_{\sigma,\sigma'} u_{j',\sigma'}(t)
\end{align}
where $\bar{\varphi}_{\sigma,\sigma'} \equiv \langle \sigma\vert \varphi\vert \sigma'\rangle$ ($\sigma,\sigma'=0,1$) are the drive-independent matrix elements in the basis formed by $\vert 0\rangle$ and $\vert 1\rangle$. Interestingly, we can use this relation to prove that the sum of all possible $|\varphi_{jj'}(t)|^2$ equals the sum of $|\bar{\varphi}_{\sigma,\sigma'}|^2$:
\begin{widetext}
\begin{align}
\label{matrix_elements}
\sum_{j,j'}  |\varphi_{jj'}(t)|^2 &= \sum_{j,j'}\sum_{\sigma,\sigma',\sigma'',\sigma'''} u_{j,\sigma}(t)u^*_{j',\sigma'}(t)u^*_{j,\sigma''}(t)u_{j',\sigma'''}(t)   \bar{\varphi}^*_{\sigma\sigma'}\bar{\varphi}_{\sigma''\sigma'''}\nonumber\\
&= \sum_{\sigma,\sigma',\sigma'',\sigma'''} \delta_{\sigma,\sigma''}\delta_{\sigma',\sigma'''}\bar{\varphi}^*_{\sigma\sigma'}\bar{\varphi}_{\sigma''\sigma'''}\quad \quad \left(\because \sum_{j}u^*_{j,\sigma}(t)u_{j,\sigma'}(t)=\delta_{\sigma,\sigma'}\right)  \nonumber \\
&= \sum_{\sigma,\sigma'}|\bar{\varphi}_{\sigma\sigma'}|^2.
\end{align}
\end{widetext}
Substituting $\varphi_{jj'}(t) = \sum_k \varphi_{jj'}^{(k)}\,e^{-ik\Omega t}$ on the left-hand side of Eq.~\eqref{matrix_elements} and averaging over one drive period gives
\begin{align}
\sum_{j,j',k} |\varphi_{jj'}^{(k)}|^2= \sum_{\sigma,\sigma'}|\bar{\varphi}_{\sigma\sigma'}|^2.
\label{vaphijjksigma}
\end{align}
Note that the right-hand side of Eq.\eqref{vaphijjksigma} is a constant, independent of drive parameters and time. Lastly, we have  
\begin{align}
\frac{1}{2}\sum_{k} |\varphi_{11}^{(k)}+ \varphi_{00}^{(k)}|^2 = \frac{1}{2}|\bar{\varphi}_{00} + \bar{\varphi}_{11}|^2,
\label{trace_remover}
\end{align}
which can be derived in a similar manner using the fact $\varphi_{11}(t) + \varphi_{00}(t) = \bar{\varphi}_{00} + \bar{\varphi}_{11}$. Subtracting Eq.~\eqref{trace_remover} from Eq.~\eqref{vaphijjksigma}, we finally arrive at
\begin{align}
\sum_k \left[2|\varphi_{01}^{(k)}|^2 + \frac{1}{2}|\varphi_{11}^{(k)}-\varphi_{00}^{(k)}|^2 \right]= 2|\bar{\varphi}_{01}|^2+\frac{1}{2}|\bar{\varphi}_{11}-\bar{\varphi}_{00}|^2.
\end{align}
Clearly, when the pure-dephasing weights $\sum_k |\varphi_{11}^{(k)}-\varphi_{00}^{(k)}|^2/2$ are suppressed, those for depolarization $\sum_k |\varphi_{01}^{(k)}|^2$ will increase. This partially explains why $T_1$ decreases at the dynamical sweet spot where pure-dephasing is suppressed. To capture the full picture of the interplay between $T_1$ and $T_\phi$, we need detailed knowledge of the noise spectra.

\bibliography{references}


\end{document}